\documentstyle[12pt]{article}
\begin{document}

\title {\bfseries{The equilibrium of dense plasma in a gravity
field}}

\author {B.V.Vasiliev}

\maketitle

\begin{center}

\itshape {Institute of Physical and Technical
Problems,Dubna,Russia,141980 $e-mail: vasiliev@main1.jinr.ru$}
\end{center}


\begin{abstract}

\bigskip

The equilibrium of dense plasma in a gravity field and problem of
a gravity-induced electric polarization in this matter are
discussed. The calculation for metals performed before shows that
both - the gravity-induced compressive strain and the
gravity-induced electric field - are inversely proportional to
their Young moduli. The calculation for high dense plasma, where
Young modulus is equal to zero, shows that there is another
effect: each cell of this plasma inside a celestial body in own
gravity field obtains the small positive electric charge. It
happens as heavy ions sag on to light electron clouds. A celestial
body stays electrically neutral as a whole, because the negative
electric charge concentrates on its surface. The gravity-induced
positive volume charge is very small, its order of magnitude
equals to $10^{-18}e$ per atom only. But it is sufficient for the
complete conterbalancing of the gravity force.

\end{abstract}

\section {The action of gravity on a metal and ultra-high density plasma}

The action of gravity on atomic substances, particularly on
metals, produces  compressive strains which is inversely
proportional to their Young moduli ${Y}$ \cite{1}. The equilibrium
equation of an atomic substance is

\begin{equation}
\gamma \vec{g}+\nabla{P}=0 \label{10}
\end{equation}

Where $\nabla P$ is the gradient of gravity-induced pressure,
$\gamma$ is the density of a metal, $\vec{g}$ is the gravity
acceleration.

The gravity-induced electric polarization in metal has often been
a subject for investigation before (see, for example \cite{2}).

The basic result of those investigations may be reduced to the
statement that gravity induces inside a metal an electric field
with an intensity of

\begin{equation}
\overrightarrow{E}\simeq\frac{nE_F}{Y}\frac{m_{i}}{e}
\overrightarrow{g},
\label{110}
\end{equation}

Where $m_{i}$ is the mass of an ion, $e$ is an electron charge,
$n$ and $E_F$ is the density and Fermi-energy of electrons.

Under conditions of the Earth, this field is so small that it is
not possible to measure it experimentally.

It is a direct consequence of the presence of an ion lattice
inside metal. This lattice is deformed under the action of
gravity, depending on its Young modulus, and then the electron gas
adapts its density to this deformation, depending on its
compressibility $nE_F$.

As under the action of ultra-high pressure, all substances
transform into high density electron-nuclear plasma. For this
state Young modulus $Y=0$ and the action of gravity on plasma
demands a special consideration.

First, let us mentally divide plasma into spherical cells. The
volume of a cell must be equal to a volume of plasma related to an
ion. The radius $r_s$ of such a spherical cell in plasma with the
mass density $\gamma$ and the density of electrons are given by

\begin{equation}
\left(\frac{\gamma}{m_{i}}\right)^{-1}=
\frac{4\pi}{3}r_{s}^{3}=\frac{Z}{n}.\label{120}
\end{equation}

Where Z and A are the charge and  atomic number of nucleus,
$m_{i}=Am_{p}$ is the mass of nucleus, $m_{p}$ is the mass of
proton,

and the density of electrons

\begin{equation}
n=\frac{3Z}{4\pi{r_{s}^{3}}}.\label{130}
\end{equation}

Now let us write down  the equilibrium condition of plasma. Here,
in nuclear subsystem, the direct interaction between nuclei is
absent. Therefore the equilibrium of nuclear subsystem of
eN-plasma (at T=0) looks like

\begin{equation}
\mu_{i}=Ze\Phi+m_{i}\Psi=const.\label{140}
\end{equation}

Where $\Psi$ is the potential of a gravitational field, $\Phi$ is
the potential of electric field.

The direct action of gravitation on electrons can be neglected due to
their small mass.
Therefore, the equilibrium condition in electron gas obtains the form

\begin{equation}
\mu_{e}=\varepsilon_{F}= \frac{p_{F}^{2}(r)}{2m}-(e-
\delta{q})\Phi=const.\label{150}
\end{equation}

By introducing  the charge $\delta q$, we take into account that
the charge of an electron cloud inside a cell can differ from Ze.
A small number of electrons can stay on the surface of a plasma
body where the electric potential is absent. As a result, the
charge of a cell, subjected to the action of the electric
potential, is effectively decreased by a small value $\delta q$.
If the radius of a star $R_0$ is approximately $10^{10} cm$, one
can expect that this mechanism gives an order of magnitude
$\frac{\delta q}{e}\approx \frac{r_s}{R_0}\approx 10^{-18}$.

The electric potential inside each cell consists both of the
potential of a considered cell $\varphi (r)$  and the potential,
which is induced by other cells $\varphi(R)$:

\begin{equation}
\Phi=\varphi (r) + \varphi (R).\label{160}
\end{equation}

The electrostatic potential of the arising field is determined by
the Gauss law

\begin{equation}
\frac{1}{r^{2}}\frac{d}{dr}\left[r^{2}\frac{d} {dr}\Phi\right]=
-4\pi\left[Ze\delta(r)-en(r)\right].\label{170}
\end{equation}

where the position of nuclei is described by the function $\delta
(r)$.

\section {The Thomas-Fermi approximation}

According to the Thomas-Fermi method, the density of electrons is
approximated by

\begin{equation}
n(r)=\frac{8\pi}{3h^{3}}p^{3}_{F}(r).\label{210}
\end{equation}

With this substitution, Eq.({\ref{170}}) is converted into a
nonlinear differential equation for $\Phi$, which for $r>0$ is
given by

\begin{equation}
\frac{1}{r^{2}}\frac{d}{dr}\left(r^{2}\frac{d}{dr}\Phi\right)=
4\pi\left[\frac{8\pi}{3h^{3}}\right]
\left[2m_{e}(\mu_{e}+(e-\delta{q})\Phi)\right]^{3/2}.\label{220}
\end{equation}

It can be simplified by introducing the following variables
\cite{3}

\begin{equation} \mu_{e}+(e-
\delta{q})\Phi=Ze^{2}{\frac{u}{r}}\label{230}
\end{equation}

and $r=ax$.

Where $a=\left(\frac{9\pi^{2}}{128Z}\right)^{1/3}a_{0}$,

and $ a_{0}=\left(\frac{\hbar^{2}}{me^{2}}\right)$ is the Bohr
radius.

With an allowance for Eq.({\ref{230}})

\begin{equation}
Ze^{2}{\frac{u}{r}}= const -\frac{m_{i}\Psi}{Z}-
\delta{q}\Phi.\label{240}
\end{equation}

then Eq.({\ref{220}}) is transformed to

\begin{equation}
\frac{d^{2}u}{dx^{2}}=\frac{u^{3/2}}{x^{1/2}}.\label{250}
\end{equation}

In terms of $u$ and $x$, the electron density within a cell is
given by

\begin{equation}
n_{x}=\frac{8\pi}{3h^{3}}p^{3}_{F}=
\frac{32Z^{2}}{9\pi^{3}a^{3}_{0}}\left(\frac{u}{x}\right)^{3/2}.
\label{270}
\end{equation}

\section {The gravity-induced electric charge of
a cell of ultra-high density plasma}

The full charge of a cell under absence of gravitation is zero.
Under influence of gravitation, the charge of the electron gas in
a cell becomes equal to

\begin{equation}
Q_{e}=4\pi{e}\int^{r_{s}}_{0}n(r)r^{2}dr=\frac{8\pi{e}}{3h^{3}}
\biggl[2m\frac{Ze^{2}}{a}\biggr]^{3/2}4\pi{a}^{3}
\int^{x_{s}}_{0}x^{2}dx\biggl[\frac{u}{x}\biggr]^{3/2}.
\label{310}
\end{equation}

Using Eq.({\ref{250}}), we obtain

\begin{eqnarray}
Q_{e}=Ze\int^{x_{s}}_{0}xdx\frac{d^{2}u}{dx^{2}}=
Ze\int^{x_{s}}_{0}dx\frac{d}{dx}\biggl[x\frac{du}{dx}-u\biggr]=
\nonumber \\=Ze\biggl[x_{s}\frac{du}{dx}\bigg|_{x_{s}}-
u(x_{s})+u(0)\biggr].\label{320} \end{eqnarray}

At $r\rightarrow0 $ the main part of electric potential is due to
nuclei alone $ \Phi \rightarrow\frac {Ze} {r} $. It means that $
u(0)\rightarrow1 $ and each cell of plasma obtains a small charge

\begin{equation}
\delta{q}=Ze\biggl[{x_{s}}\frac{du}{dx}\bigg|_{x_{s}}-u(x_{s})\biggr]=
Zex_s^2\biggl[\frac{d}{dx}\biggl(\frac{u}{x}\biggr)\biggr]_{x_s}
\label{322}
\end{equation}

If a radius of a cell can be a function of some parameters,
Eq.({\ref{322}}) transforms to

\begin{equation}
\delta{q}=Ze{r_{s}}^2\frac{d}{dr_s}\biggl(\frac{u(r_{s})}{r_s}\biggr).
\label{334}
\end{equation}

When the charge of a cell depends on its location  inside a star

\begin{equation}
\delta{q}=Zer_{s}^2\biggl[\frac{d}{dR}
\biggl(\frac{u(r_s)}{r_s}\biggr)\biggr]\biggl[\frac{dR}{dr_{s}}\biggr].
\label{336}
\end{equation}

Because inside a spherically symmetric star the gravity
acceleration is

\begin{equation}
\vec{g}=-\frac{d\Psi}{dR}\frac{\overrightarrow{R}}{R} \label{338}
\end{equation}

and the electric field intensity is

\begin{equation}
\vec{E}=-\frac{d\Phi}{dR}\frac{\overrightarrow{R}}{R}\label{340}
\end{equation}

from Eq.({\ref{336}})

\begin{equation}
\frac{dr_{s}}{dR}\frac{\overrightarrow{R}}{R}=
\frac{r_{s}^2}{e\delta{q}}\biggl[\frac{m_{i}}{Z}\vec{g}
+\delta{q}\vec{E}\biggr].\label{350}
\end{equation}

This equation has the following solution

\begin{equation}
\frac{dr_{s}}{dR}=0\label{360}
\end{equation}

and

\begin{equation}
\frac{m_{i}}{Z}\vec{g}+\delta{q}\vec{E}=0.\label{370}
\end{equation}

As

\begin{equation}
div{\vec{g}}=-4\pi{G}{n}m_{i}\label{372}
\end{equation}

and

\begin{equation}
div{\vec{E}}=4\pi{n}\delta{q}\label{375}
\end{equation}

the gravity-induced electric charge in a cell is very small

\begin{equation}
\delta{q}=\sqrt{G}\frac{m_{i}}{Z}\simeq{10^{-18}e},\label{380}
\end{equation}

where $G$ is the gravitation constant.

In plasma the equilibrium value of the electric field on the
nuclei is approximately equal to $g m_i/e$. This value is, like
that in a metal, very small. But there is one additional effect in
plasma. Simultaneously with the confinement of nuclei in
equilibrium, each cell obtains an extremely small electric charge.
However, because the size of bodies is very large, the electric
field intensity may be very large as well

\begin{equation}
\vec{E}=\frac{\vec{g}}{\sqrt{G}}\label{390}
\end{equation}

Thus, a celestial body is electrically neutral as a whole, because
the positive volume charge is concentrated inside the charged core
and the negative electric charge exists on its surface and so one
can infer gravity-induced electric polarization of a body.

In accordance with Eqs.({\ref{360}})-({\ref{370}}) the action of
gravity here is completely compensated by the electric force and
the pressure gradient is absent. Instead of Eq.({\ref{10}}) which
is correct for atomic substances, these equations describe the
equilibrium into celestial bodies which consist of high dense
plasma.

\end{document}